# Effects of Stirring Time on Formation of Microplastics Fragmented from Photo-aged Polypropylene

*Kazuya Haremaki[a], Takumitsu Kida[b], Yusuke Koide[a], Takashi Uneyama[a], Yuichi Masubuchi[a], Takato Ishida[a*]*

[a] Department of Materials Physics, Nagoya University, Furo-cho, Chikusa, Nagoya 464-8603, Japan

[b] Department of Materials Chemistry, Faculty of Engineering, The University of Shiga Prefecture, 2500, Hassaka, Hikone, 522-8533, Japan

* E-mail: ishida@mp.pse.nagoya-u.ac.jp

**RUNNING HEAD**

Effect of Stirring Time on Formation of Microplastics

**GRAPHICAL ABSTRACT**

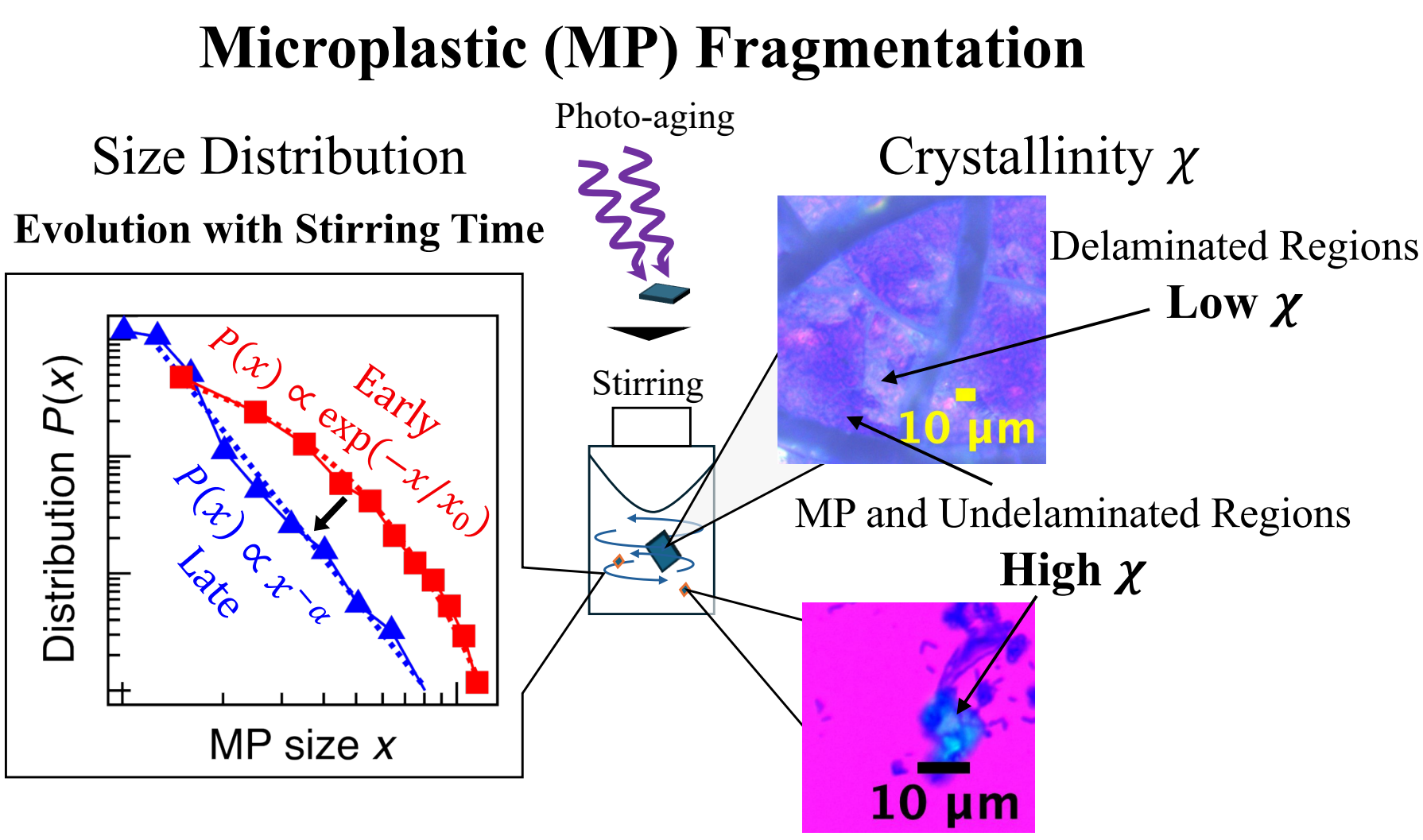

## ABSTRACT

This paper examines the evolution of microplastic (MP) size distributions fragmented from photo-aged polypropylene (PP) in stirred water. PP specimens fragmented into MPs with their size of 1–30 μm after UV irradiation and stirring in laboratory settings. These laboratory-fragmented MPs were dispersed into the water during the stirring process. A series of MP size distributions was analyzed from optical microscope images of obtained MPs. The MP size distribution was described by an exponential function in the short stirring time domain, whereas it changed to a power-law function as the stirring time increased. The fragmentation rate of MPs and nanoplastics (NPs) decreased with increasing stirring time. The obtained MP exhibited higher crystallinity than the photo-aged PP specimen after stirring. This result implies that MP fragmentation, as observed under controlled laboratory conditions, is related with the chemi-crystallization of PP.

## 1. INTRODUCTION

Microplastics (MPs) are small plastic particles typically defined as ranging in size from 1 μm to 5 mm [1]. MPs are known to originate from the fragmentation of plastic debris. They have been detected in a wide variety of environments, including soils, river water, the atmosphere and even within biological systems such as the human body [2–9]. Among these, the marine environment is considered the most significant reservoir of MPs. The presence of MPs in the marine environment has become a serious problem today. The size of MPs affects transport and fragmentation process in the ocean[10,11]. MPs of certain sizes may pose greater ecological risks, highlighting the importance of understanding and mitigating their fragmentation. Therefore, elucidating the mechanisms of MP fragmentation and identifying key factors that induce fragmentation are crucial.

Halle et al. [11] proposed a widely accepted scenario for MP fragmentation from plastic waste in marine environments: (i) embrittlement caused by photo-oxidative degradation, followed by (ii) MP formation under mechanical stress. In the polyolefin family, which represent a major source of MPs, oxidation leads to chain scission primarily in the amorphous region. This process also induces chemi-crystallization, resulting in embrittlement [12–14]. In severely aged samples, it has been known since the 1980s that chemi-crystallization induces internal stress distribution, and the stress inhomogeneity is released through the spontaneous formation of orthogonal surface cracks [15,16]. MPs are subsequently formed when mechanical stress fragments the embrittled surface and propagates along the crack edges [17]. In the marine environment, marine plastic debris experiences mechanical stresses arising from various external stimuli, including waves, wind, abrasion by sand, and interactions with marine organisms [18,19].

Several studies have reported statistical analyses of the size distribution of MPs collected from both environments and laboratory tests. These studies commonly indicate that the size distribution follows a power-law distribution [11,17,20–26]. A simple analytical model has been proposed to explain this power-law behavior. It claims that the power-law with the exponent of 3 is realized, assuming that cubic elements undergo a cascading fragmentation process [23]. However, other previous studies showed that the power-law exponent remains uncertain [20–22], likely due to some differences in fragmentation conditions and the observed MP size ranges across studies. According to a simple cascading model, the exponent varies depending on the dimensionality of the objects (fibers, films or bulk particles), and ranges from 1 to 3 [27]. Another possible reason for the inconsistency is the lack of studies that consider the time evolution of MP fragmentation behavior.

In this study, photo-aged polypropylene (PP) specimens with a fixed oxidative aging treatment were prepared and stirred in water to fragment MPs under controlled laboratory conditions, and the time evolution of the laboratory-fragmented MP size distribution was investigated. The laboratory fragmentation test setup was designed such that relatively small, micrometer-scale MPs were primarily generated through delamination from the embrittled surface of the photo-aged PP specimens. Throughout this paper, we refer to these laboratory-fragmented MPs simply as “MPs”.

The fragmented specimens were observed by optical microscopy, Raman spectroscopy, infrared (IR) spectroscopy, and differential scanning calorimetry (DSC). The results showed that (i) the size distribution changed from exponential to power-law distribution as the stirring time increased and that (ii) high crystallinity regions are preferentially fragmented into MPs. Details are shown below.

## 2. EXPERIMENTAL

### 2.1 Materials and Specimens Preparation

The isotactic polypropylene (PP) used in this study was obtained from Sigma-Aldrich (weight average molecular weight $M_w = 3.5 \times 10^5$, and polydispersity index $M_w/M_n = 3.6$). The supplied pellets of PP were placed between two 100 μm-thick and 150 mm × 150 mm aluminum sheets coated with polyimide films, together with a 1 mm-thick and 150 mm × 150 mm aluminum spacer featuring a 100 mm × 100 mm square cut-out. PP was molded at 200 ºC and 15 MPa for 15 minutes by a test press (MP-2FH, Toyo-Seiki). The pressed sheets were then placed between two 5 mm-thick and 200 mm × 200 mm stainless steel plates and cooled gradually to room temperature (23 °C) for approximately one minutes. The obtained sheets were cut into 10 mm × 10 mm specimens. The initial crystallinity of the PP specimen, evaluated by DSC, was found to be 38% by weight. The enthalpy of fusion of 209 J/g for the fully crystalline PP was taken for the crystallinity calculation [28].

### 2.2 Photo-aging

The photo-aging treatment was conducted under an OPM2-502XQ xenon lamp (USHIO Inc., Japan) while the 10 mm × 10 mm specimens were placed on a house-made hot plate of which temperature was controlled to be 70 ºC. The mirror mounted on the lamp ensured horizontal irradiation for the specimen. The average UV radiation flux density was 6.54 mW/cm$^2$ at 365 nm, measured by UIT-250 accumulated UV meter (USHIO Inc., Japan), with UVD-S365 photodetector (USHIO Inc., Japan).

### 2.3 Fragmentation

A single photo-aged specimen was placed in a 30 mL screw tube vial (60 mm height, 30 mm diameter) containing 15 mL of distilled water and stirred using a VTX-3000L vortex mixer (LMS Co. Ltd., Japan) at 500 rpm for up to 400 h to fragment MPs. Throughout the stirring process, the remaining specimen was periodically removed from the vial and transferred to a new vial containing fresh 15 mL deionized water at 5, 10, 20, 50, 100, 200, and 400 h of stirring. After transferring the specimens, the water remaining in the vial, including MPs, at each stirring period (0–5 h, 5–10 h, 10–20 h, 20–50 h, 50–100 h, 100–200 h, and 200–400 h) was obtained. Because the MP fragmentation rate would decrease with time, the period between transfers was increased as the stirring time increased. In this study, we focus on systems in which the specimen surface is well-aged, and MPs delaminated from the embrittled surface during stirring are the main target of observation.

**2.4 Characterization of The Remaining PP Specimen**

At each transfer between vials, the PP specimen was analyzed as follows. The surface was observed by a polarizing optical microscope (BX53, OLYMPUS, Japan) that was equipped with a crossed polarizer and a retardation plate. Reflection illumination was employed for observations of the specimen photo-aged and stirred for 400 h after photo-aging, with a spatial resolution of the imaging was 0.22 μm/pixel. Transmission illumination was employed for observations of the unaged specimen sliced by a microtome (THK, KENIS Ltd., Japan), with a spatial resolution of the imaging was 0.11 μm/pixel.

The infrared (IR) spectra were taken by Nicolet iS10 FT-IR spectrometer (Thermo Fisher Scientific, USA). Attenuated total reflection (ATR) measurements were performed with an ATR accessory (ATR ITX base, Thermo Fisher Scientific, USA) and a diamond crystal (ATR ID7/ITX noncoated diamond crystal, Specac Ltd, Britain). All IR spectra were collected over a range of 4000–525 $cm^{-1}$ with a resolution of 1 $cm^{-1}$, and each spectrum was averaged over 64 scans.

DSC analysis was conducted with a Discovery DSC 25 (TA Instruments, USA) over a temperature range from 50 ºC to 200 ºC at a heating rate of 10 ºC/min under a nitrogen atmosphere. The samples were sealed in a closed pan to prevent the loss of volatile components.

After photo-aging and stirring, Raman spectroscopy of PP specimens and MP was conducted using inVia Reflex Raman micrometer (Renishawm, UK) with a green laser (wavelength: 532 nm, power: 50 mW) and an objective lens ×100. Each Raman spectrum was accumulated 32 times with an exposure time of 2 s under a non-polarized condition by inserting a 1/4 wavelength filter in the incident-light path. All Raman spectra were collected over a range of 1800–5 $cm^{-1}$ with a resolution of 2 $cm^{-1}$.

Mapping measurement of the orientation of unaged PP specimen was conducted using inVia Reflex Raman microscope (Renishaw, UK) with an objective lens ×100 on a selected area of 150 μm×150 μm with a step size of 3 μm. Each Raman spectrum was obtained over a range of 1800–5 $cm^{-1}$ with a resolution of 2 $cm^{-1}$ under *hh* polarization, where the polarization direction of the incident light was parallel to the scattered light, with an exposure time of 0.2 s.

### 2.5 MP Size Distribution Measurements

After the removal of the specimen during the fragmentation process, to obtain MP size distribution at each stirring period, the remaining water containing MPs was processed as follows. The liquid was partially dried in AVO-310NB vacuum oven (AS ONE Co., Japan) at 70 °C in order to increase the MPs concentration. A single drop of the concentrated MPs dispersion was sampled with a glass rod and transferred to a slide glass. Afterward, the water on the slide glass was fully evaporated. Polarized optical microscope images of MPs on the slide glass were taken by the microscope mentioned in section 2.4, under a transmission illumination. The spatial resolution of the imaging was 0.11 μm/pixel. The acquired images were processed by ImageJ2 (version 2.14.0). The hue channel was binarized, and Feret's diameters of MP particles were measured. The Feret's diameter corresponds to the maximum size measured by the caliper. The MP size, defined as the Feret's diameter, was measured for at least 1000 MP particles. Then, the MP size distributions were organized as histograms. The bin size is 1 μm for linear and semi-logarithmic scales and logarithmically spaced for the logarithmic scale. The MP particles smaller than 1 μm were excluded from the size distributions.

### 2.6 MP Fraction Measurements

Apart from the MP size distribution measurement, the amount of fragmented MPs and nanoplastics (NPs) at each stirring period was determined as follows. The concentrated MP dispersion, prepared in section 2.5, was completely evaporated in an AVO-310NB vacuum oven (AS ONE Co., Japan) at 70 °C. The remaining MPs and NPs in the vial were then diluted with about 3 mL acetone as a solvent. The acetone solution containing MPs and NPs was transferred to a 5 mL glass syringe and dripped at a constant rate of 2.00 mL/h into an aluminum DSC pan with an MSP-1D syringe pump (AS ONE Co., Japan). The temperature of the DSC pan was kept at 50 ºC by the house-made hot plate in order to quickly evaporate the solvent. To collect MPs remaining in the vial, acetone was added to the vial again and the same procedure was iterated more than three times to reduce the loss of MPs as much as possible. In order to evaporate the solvent completely, the DSC pan was dried in an AVO-310NB vacuum oven (AS ONE Co., Japan) at 60 °C for 12 h.

The change in weight of the pan is regarded as the weight of fragmented MPs and NPs. The accumulated weight of fragmented MPs and NPs was calculated as the sum of the measured weights:

$$W_{\mathrm{frag},n} = \sum_{i=1}^{n} w_{\mathrm{frag},i} \, . \tag{1}$$

Here, $w_{\mathrm{frag},i}$ represents the weight of MPs and NPs fragmented during the $i$-th stirring period, and $W_{\mathrm{frag},i}$ is the accumulated weight up to the $i$-th stirring period. The average fragmentation rate of MPs and NPs during the $i$-th stirring period is defined as follows:

$$R_{\mathrm{frag},i} = \frac{w_{\mathrm{frag},i}}{t_i - t_{i-1}} \, , \tag{2}$$

where $t_i$ is the end of the $i$-th stirring period (i.e., 5 h, 10 h, 20 h, 50 h, 100 h, 200 h, and 400 h).

## 3. RESULTS

### 3.1 Observation of MPs by Polarized Optical Microscopy

Figure 1 shows polarized optical microscope images of artificially fragmented MPs at each stirring period (20–50 h, 50–100 h, 100–200 h, and 200–400 h). Some waxy residues, not identified as MPs, were observed during the early stage (up to 20 h). The size of the MPs ranges from larger than 1 μm to smaller than 30 μm, with various shapes, consistent with the earlier studies [29–33]. The size of MPs fragmented before 200 h of stirring did not appear to differ significantly from each other, whereas larger MPs were fragmented after 200 h of stirring.

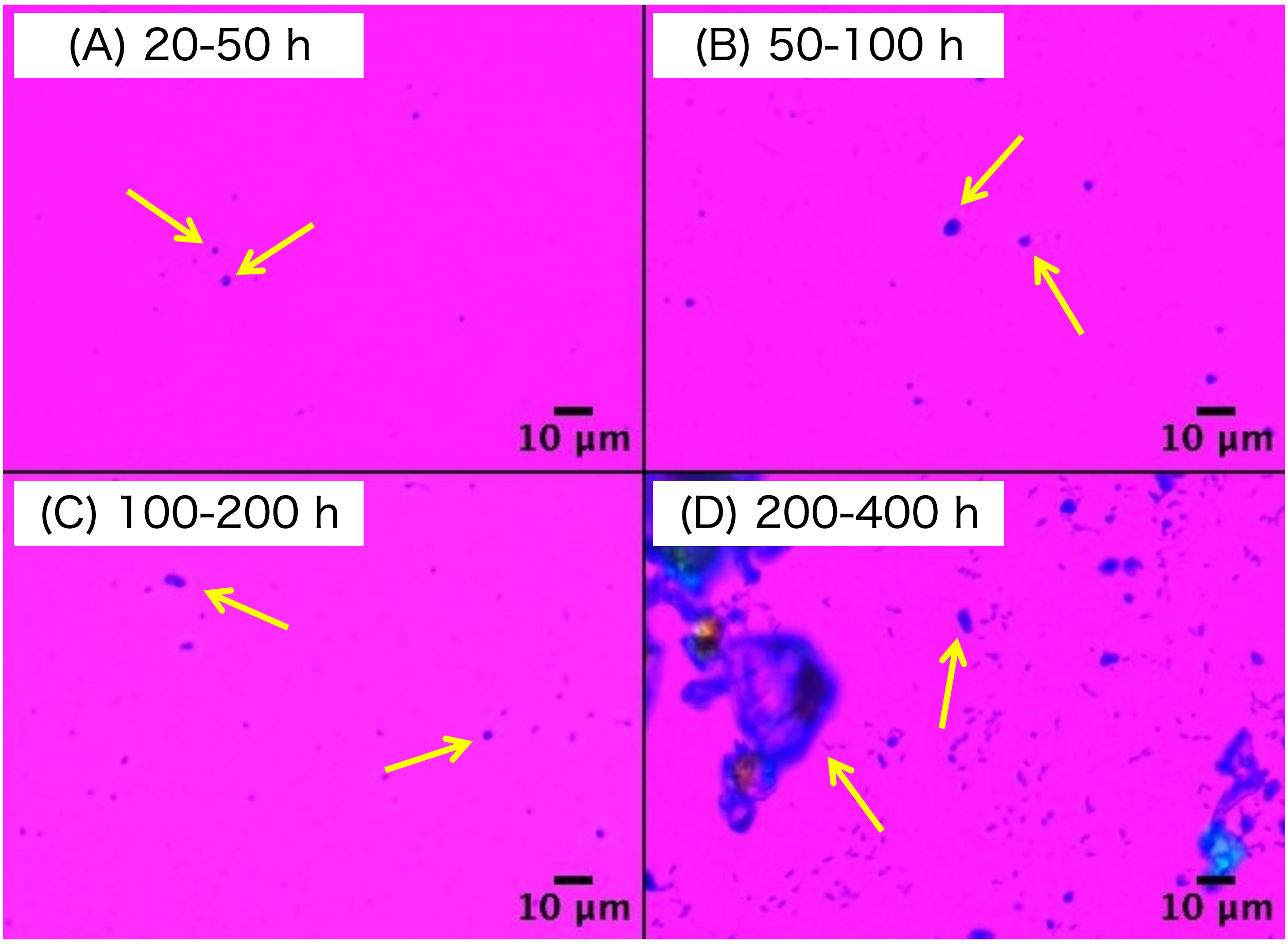

Figure 1 (A-D) Polarized optical microscope images of MPs fragmented from photo-aged PP specimens after more than 20 h of stirring. Typical MPs are indicated by the yellow arrows.

### 3.2 Evolution of MP Size Distribution

Figure 2 shows the size distributions of MPs at each stirring period (20–50 h, 50–100 h, 100–200 h, and 200–400 h) plotted in linear, semi-logarithmic, and double-logarithmic manners. MPs were not observed by the optical microscope at the early stage (less than about 20 h of stirring). Thus, only the data after 20 h of stirring are shown in Figure 2. Although the number of MPs visible in the fields of view shown in Figure 1(A–C) is limited, the MP size distributions presented in Figure 2 were obtained by surveying a much broader area, in which more than 1,000 MP particles were identified at each stirring condition. The functional form of the MP size distributions qualitatively changed with the stirring time. The MP size distributions for 20–50 h and 50–100 h exhibited an exponential dependence $P(x) \propto \exp(-x/x_0)$. Here, $x$ is the MP size, $P(x)$ is the distribution, and $x_0$ is a constant. $x_0$ can be interpreted as a characteristic MP size. In contrast, the MP size distributions for 100–200 h and 200–400 h exhibited a power-law dependence $P(x) \propto x^{-\alpha}$, where $\alpha$ is the exponent. Unlike the exponential distribution, the power-law distribution has no characteristic size. The power-law exponents observed in Figure 2C are 3.6 for 100-200 h and 3.5 for 200-400 h, which are close to but somewhat larger than the exponent of 3 expected for a purely random 3D fragmentation model [23]. To discuss the experimental uncertainty in the measurement of MP size distributions, we present the results from independent experimental runs in Figure S1. While some variation is observed in the distributions, Figure S1 demonstrates that the qualitative scenario—namely, the transition of the distribution shape from exponential (in the 50–100 h range) to power-law (in the 100–200 h range)—remains robust across different runs.

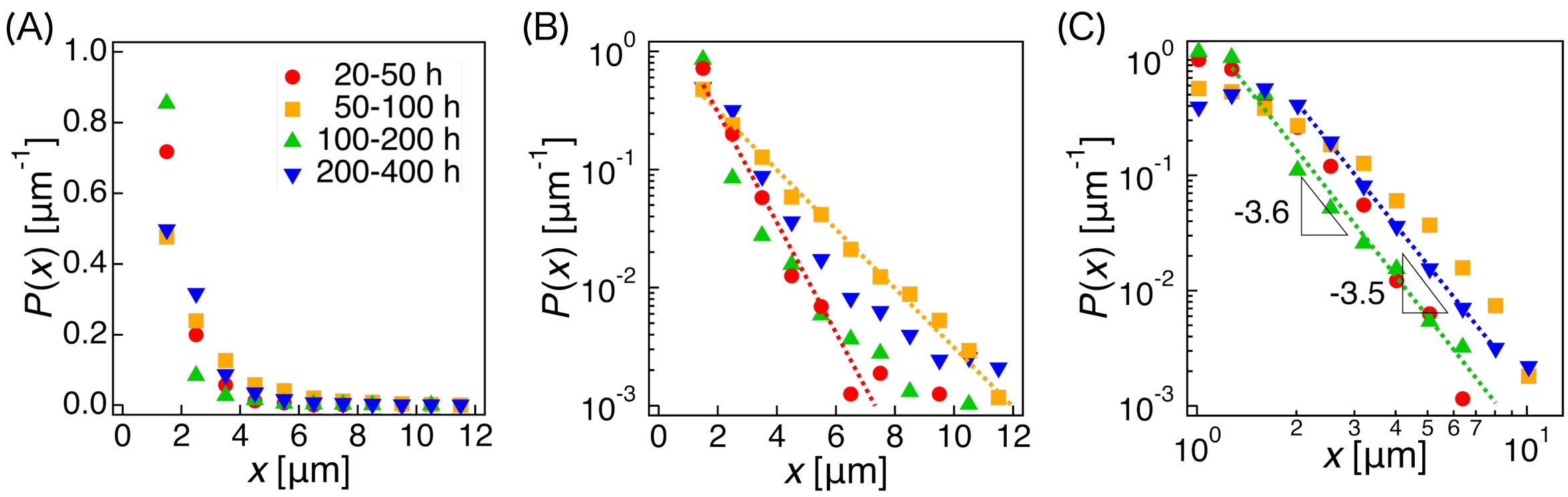

Figure 2 MP size distributions at each stirring period plotted on (A) linear, (B) semi-logarithmic, and (C) logarithmic scales. The MP size distributions for 20–50 h and 50–100 h exhibited an exponential dependence, $P(x) \propto \exp(-x/x_0)$ (lines in (B)), while the MP size distributions for 100–200 h and 200–400 h exhibited a power-law dependence, $P(x) \propto x^{-\alpha}$ (lines in (C)). The characteristic MP sizes $x_0$ for 20–50 h and 50–100 h were 0.93 μm and 1.7 μm, respectively. The power-law exponents $\alpha$ for 100–200 h and 200–400 h were 3.6 and 3.5, respectively. The bin size is 1 μm for linear and semi-logarithmic scales and logarithmically spaced for the logarithmic scale.

### 3.3 Fragmentation Rate of MPs and NPs

Figure 3 shows the average fragmentation rate ($R_{\mathrm{frag},i}$) and the accumulated weight ($W_{\mathrm{frag},n}$) of MPs and NPs as functions of stirring time. The average fragmentation rate decreased sharply until 20 h of stirring, reached a minimum at 20 h, and then slightly increased. At the longer stirring time, it started to decrease again. Here, note that $R_{\mathrm{frag},i}$ is the average value in the $i$-th stirring period, and the stirring period is not constant. The local minimum at $t_i$ = 20 h may not be accurate, and it is not discussed further in what follows. While the local minimum is not consistently observed across different experimental runs (Figure S2), the overall trend suggests a two-step fragmentation process: initially, hydrophilic soluble products generated by photo-aging are released (which are likely too small to be detected by optical microscopy), followed by the gradual release of microplastics. After 400 h of stirring, about 1.4 mg MPs were fragmented from the specimen. The weight of unaged specimens was about 0.8 g, indicating that 1/500 of the specimen fragmented into MPs.

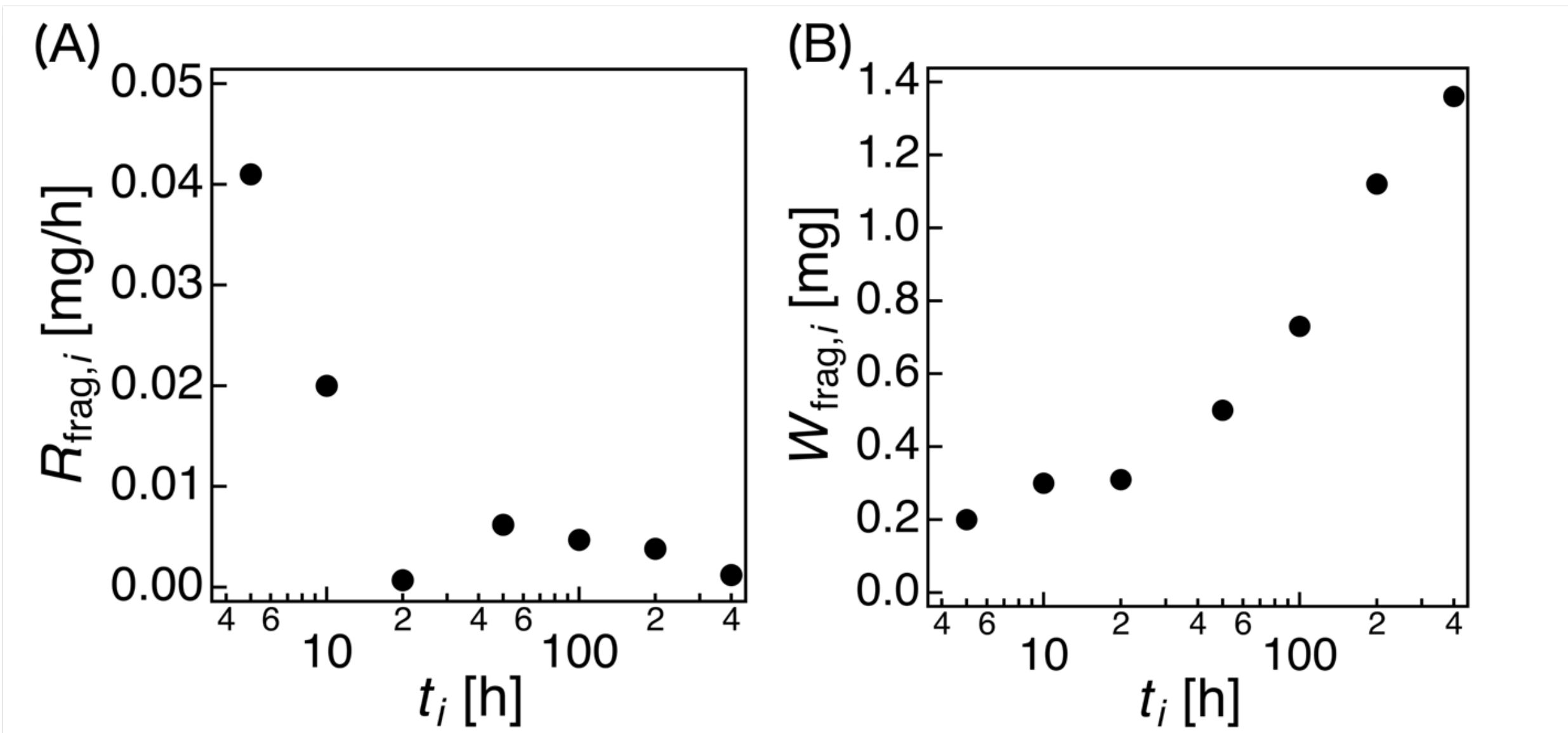

Figure 3 Time evolution of (A) the average fragmentation rate of MPs and NPs and (B) the accumulated weight of fragmented MPs and NPs. $t_i$ is end of the $i$-th stirring period.

### 3.4 Observation of Remaining PP Specimen Surface

Optical microscope images of the photo-aged PP specimen surface after different stirring times are shown in Figure 4. Even without stirring, surface embrittlement occurred during the photo-aging process, and spontaneous cracks could be observed on the specimen surface (Figure 4A), as reported earlier [15,16]. The distance between neighboring cracks was about 100–200 μm [33,34]. Some cracks were connected in a pattern resembling a capital letter "T" known as the formation of T-junction [35,36], consistent with chemi-crystallization-induced crack propagation [25,26]. Surrounded by these cracks, island-like regions on the surface of the photo-aged specimen were formed. For the stirring time of 5–200 h, in addition to the island-like regions, some narrow cracks were observed inside them. These narrow cracks were probably formed by the mechanical stresses (See Figures 4B–4G). According to the results in Figure 3, traces of MP fragmentation are expected to be observed. Indeed, compared with Figure 4A, edges formed by cracks appear more obscure. After the photo-aged specimen was stirred for 400 h (Figure 4H), mesh-like cracks and some spots with the fragmentation of MPs were observed inside the island-like regions. Also, warp along cracks was observed in Figure 4B–4H.

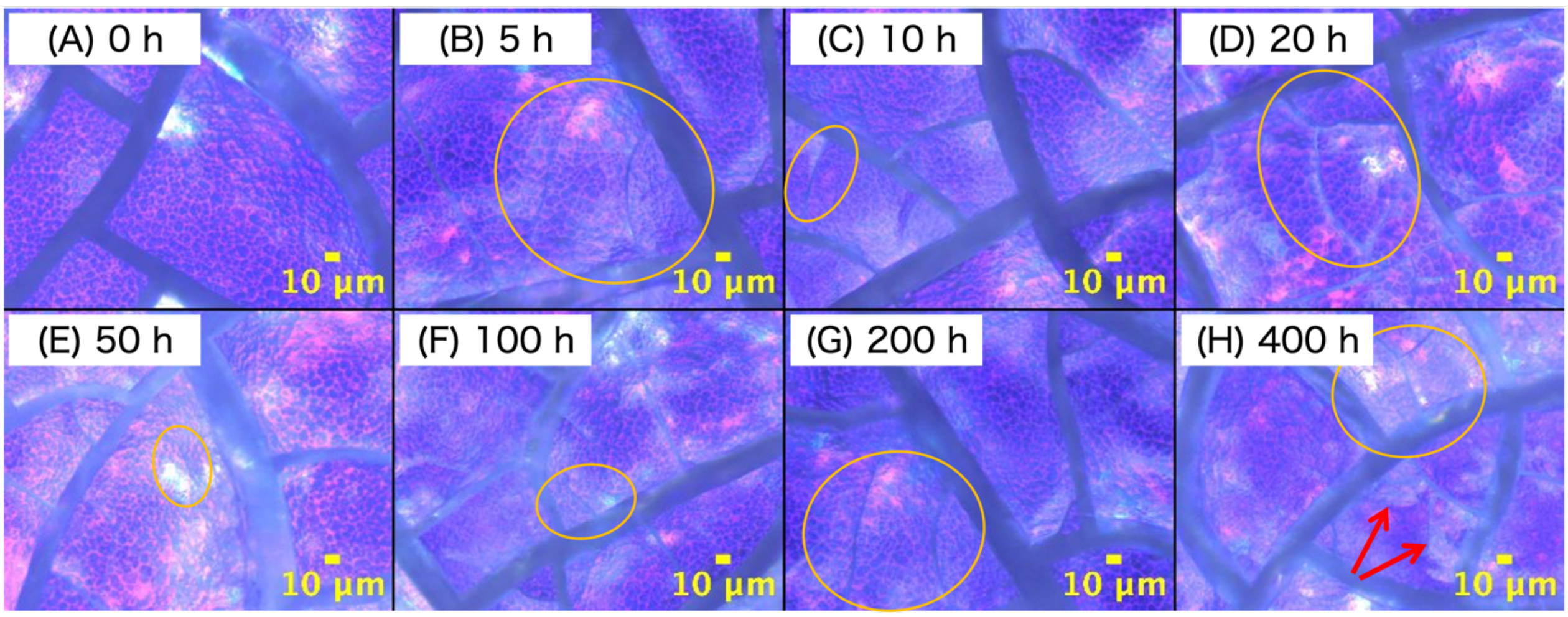

Figure 4 (A-H) Polarized optical microscope images of specimen surface after stirring for 0 h to 400 h. The locations where MPs fragmented in Figure 4H are highlighted by the red arrows. Circles

show cracks in the island-like regions.

Figure 5 shows a mapping image of the orientation parameter $R$ and a polarized optical microscope image of the unaged PP specimen. The intensity ratio of the Raman band at 973 and 998 $cm^{-1}$ is used as an orientation parameter $R = A_{973}/A_{998}$. The peaks at 973 $cm^{-1}$ and 998 $cm^{-1}$ correspond to the asymmetric C-C stretching mode of the skeletal backbones and the rocking mode of the $CH_3$ lateral alkyl groups, respectively. The higher value of $R$ indicates that the crystalline chains orient to the polarization direction [37]. Clear orientational anisotropy was observed in both the R mapping and the polarized optical image. Spherulitic structures, approximately 30 μm in size, are relatively well-defined in Figure 5(A) and (B).

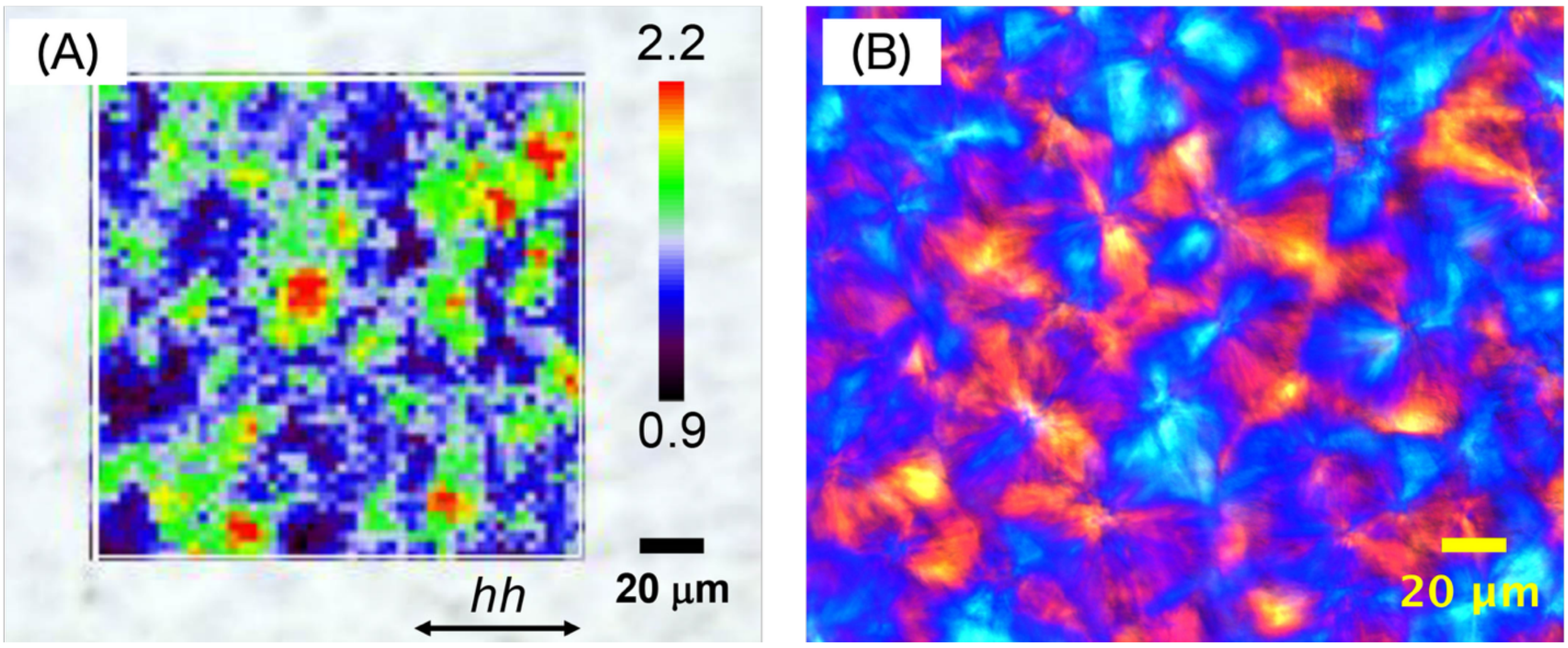

Figure 5 (A) Mapping image of the orientation parameter $R$ and (B) polarized optical microscope image of the unaged PP specimen.

### 3.5 Physico-Chemical Analysis in The Process of MP Fragmentation

The IR spectra of the PP specimens unaged, photo-aged, and stirred for 400 h after photo-aging are shown in Figure 6. IR spectra were normalized by the area of peak at 1458 $cm^{-1}$, which corresponds to the asymmetric $CH_3$ bending mode [38]. Upon photo-aging, the absorbance in the 1800–1700 $cm^{-1}$ region increases significantly. These peaks in this region are attributed to C=O vibrations, and the change is associated with the formation of end-group carboxylic acids at 1712 $cm^{-1}$ and ester structures at 1735 $cm^{-1}$ [39]. These are low-molecular-weight oxygenated products. In the IR spectrum obtained after 400 h of stirring, the absorbance for the C=O peaks decreased, reflecting the dissolution of photo-oxidized products as soluble components into water.

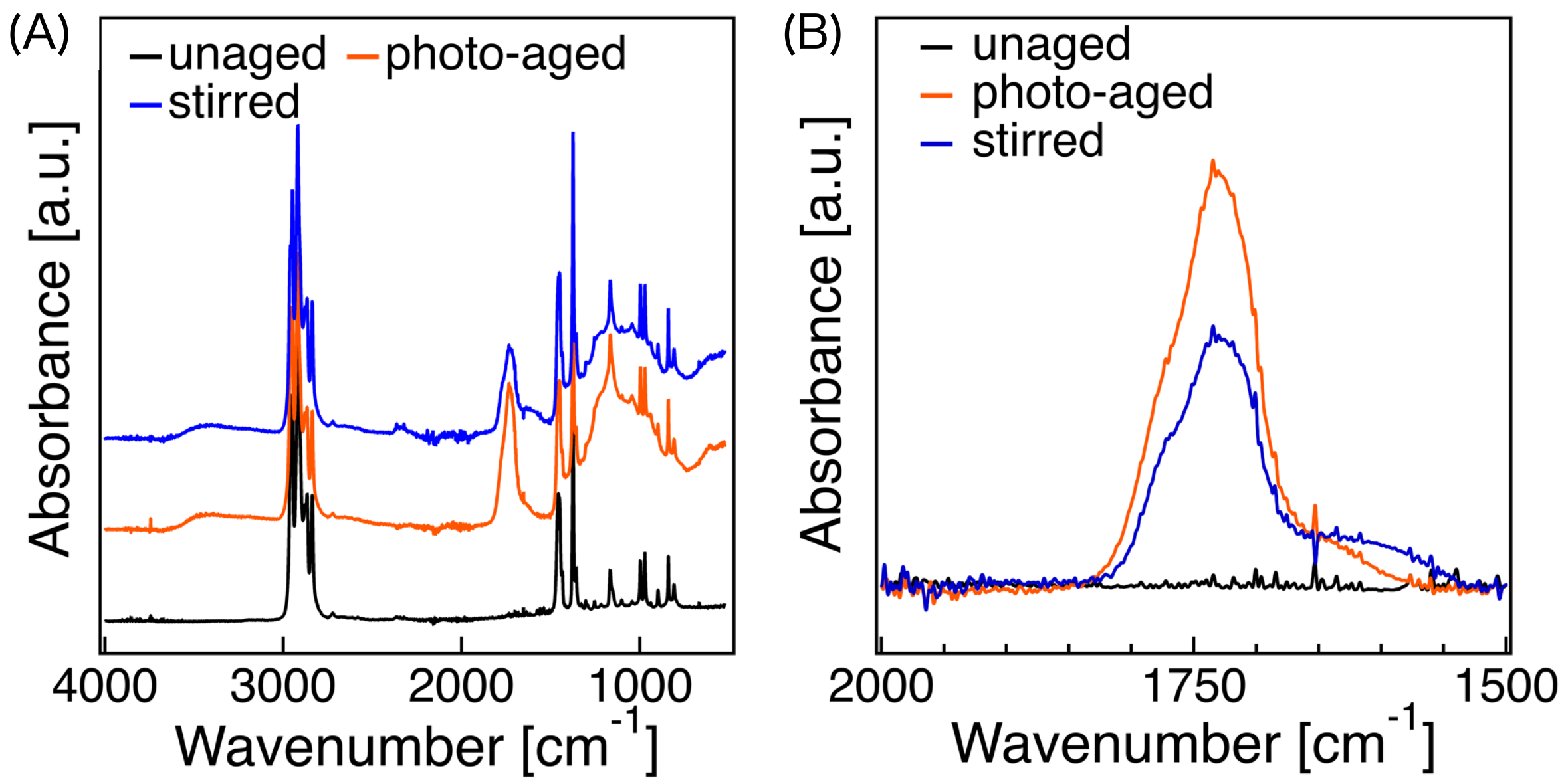

Figure 6 IR spectra of the specimen unaged, photo-aged, and stirred for 400 h after photo-aging in the regions of (A) 4000–525 $cm^{-1}$ and (B) 2000–1500 $cm^{-1}$ normalized by the area of the peak at 1458 $cm^{-1}$. The IR spectra in (B) are corrected for baseline.

Figure 7 shows the DSC traces of the PP specimens unaged, photo-aged, and stirred for 400 h after photo-aging. One can see a shift of the melting peak to the lower temperature region induced by the photo-aging. In addition, the peak shape changes, implying multiple overlapping peaks. Such behavior triggered by photo-aging has been frequently observed in earlier studies [39–41]. This result suggests the formation of low-$T_m$ (i.e., smaller) crystals due to the change in crystalline structures. Photo-aging induces chain scission, which subsequently triggers chemi-crystallization. These processes likely lead to damage of the crystalline lamellae and fragmentation of the crystalline regions. The enthalpies of fusion of the PP specimens unaged, photo-aged, and stirred for 400 h after photo-aging were 80 J/g, 86 J/g, and 84 J/g, respectively. The crystallinity clearly increases upon photo-aging, most likely reflecting the contribution of chemi-crystallization. In contrast, the stirring did not significantly alter the DSC curve and the crystallinity.

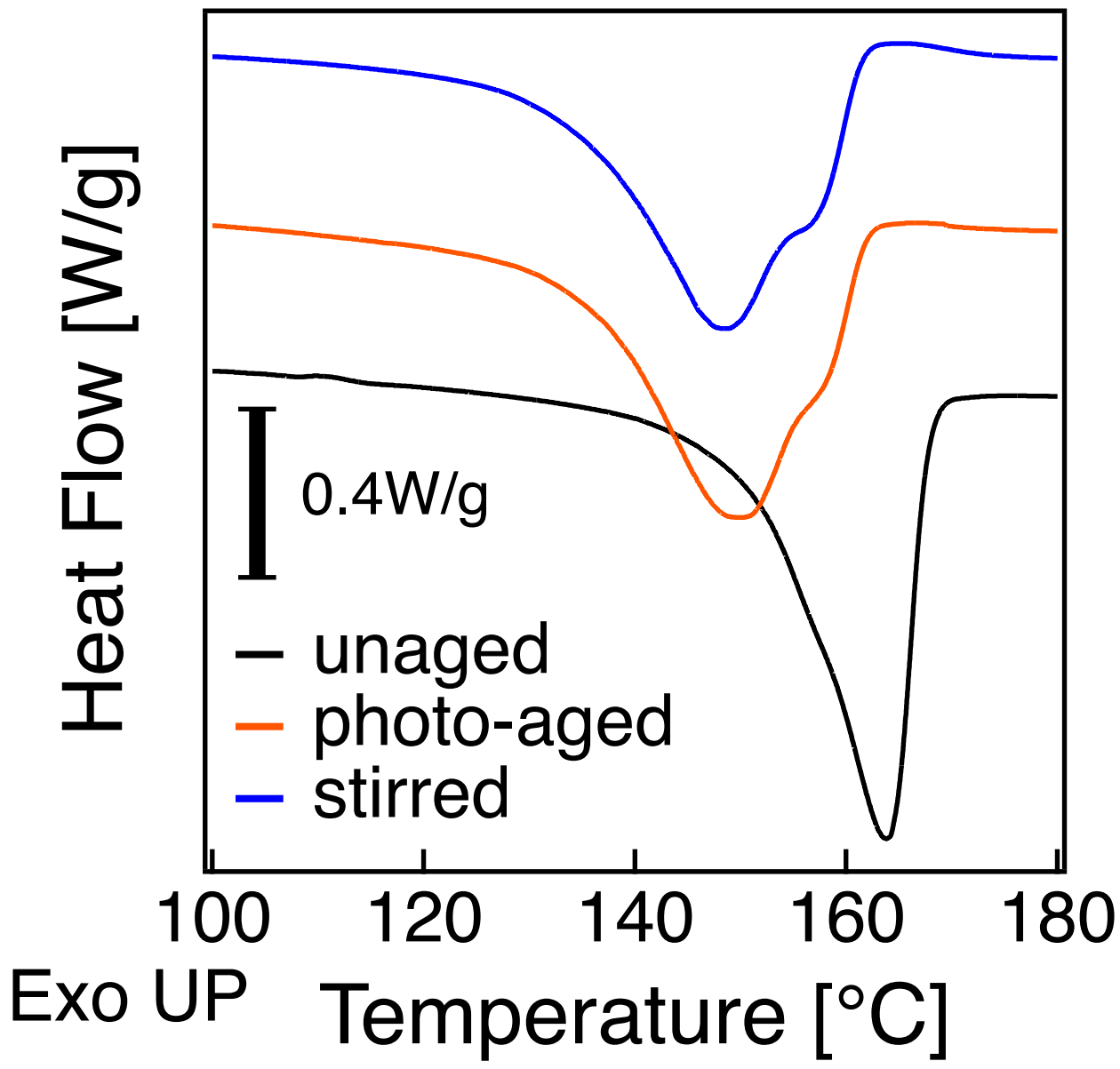

Figure 7 DSC traces of the PP specimens unaged, photo-aged, and stirred for 400 h after photo-aging.

Figure 8 shows the Raman spectra at the spots in the PP specimens unaged, photo-aged, and stirred for 400 h after photo-aging. For comparison, the Raman spectrum of a MP is also shown. All Raman spectra were normalized by the area of the peak at 1440 cm$^{-1}$, which serves as an internal standard insensitive to photo-aging [41]. The peak at 810 cm$^{-1}$ corresponds to consecutive helical C–C backbone chains with more than 12 monomers terminated by –$CH_3$, while the peak at 830 cm$^{-1}$ is attributed to amorphous backbone chains. The peaks at 830 cm$^{-1}$ and 841 cm$^{-1}$ are associated with amorphous and short helical backbone chains, respectively [42]. The Raman spectra were fitted to a sum of three Voigt functions (See Figure S3 in the Supporting Information), and peak areas of three peaks were calculated. Then the crystallinities of specimens and a MP were calculated as:

$$\chi = \frac{A_{810}}{A_{810} + A_{830} + A_{841}}. \tag{3}$$

Here, $A_{810}$, $A_{830}$, and $A_{841}$ are the peak areas for the peaks located around 810 cm$^{-1}$, 830 cm$^{-1}$, and 841 cm$^{-1}$, respectively. It was assumed that the chains in crystalline regions have sufficiently long helical sequences. The calculated crystallinities are shown in Figure 8B. The photo-aged PP specimen exhibits higher crystallinity than the unaged specimen, implying chemi-crystallization, consistent with the DSC (Figure 7). The high crystallinity of the MP and the low crystallinity of the remaining specimen imply that high crystallinity regions are preferentially fragmented into MPs. In addition, the crystallinity remarkably differs spot to spot for the specimens stirred for 400 h after photo-aging. The spot where MP fragmentation occurred exhibited lower crystallinity than that without the fragmentation.

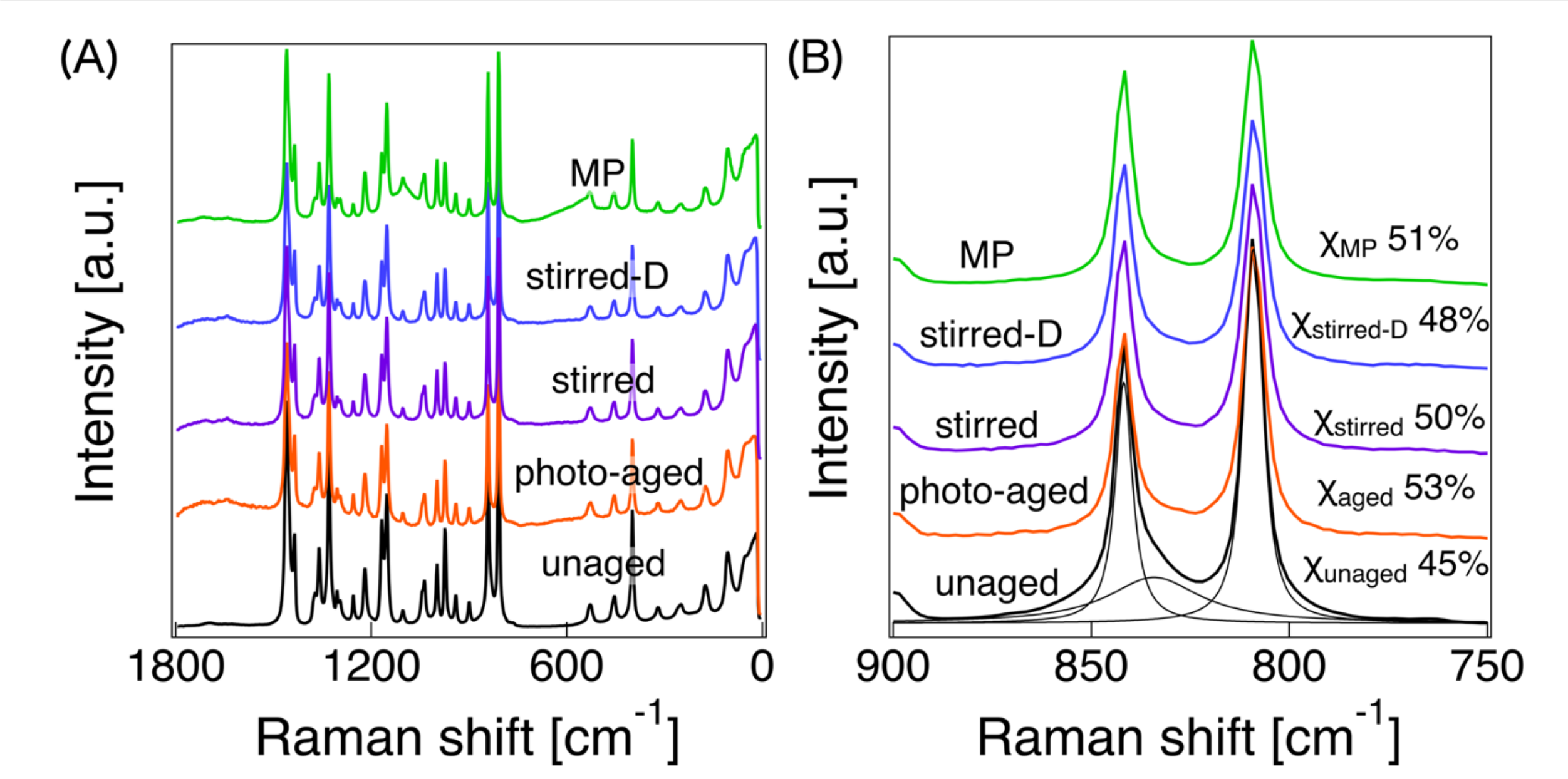

Figure 8 Raman spectra of an unaged specimen, a photo-aged specimen, several spots of the stirred specimen for 400 h after photo-aging, and MP fragmented after photo-aging followed by 400 h of stirring, in the regions of (A) 1800–0 $cm^{-1}$ and (B) 900–750 $cm^{-1}$normalized by the area of the peak at 1440 $cm^{-1}$. Spots of the stirred specimen with and without the fragmentation of MPs were labeled as “stirred-D” and “stirred”, respectively. The crystallinity at each spot is shown in (B).

## 4. DISCUSSION

In this section, from the experimental results shown above, let us discuss the formation process of laboratory-fragmented MPs under stirring. Just after photo-aging (before stirring), deep cracks appear on the surface of the PP specimen (Figure 4A) due to chemi-crystallization, which causes embrittlement of the specimen. As a side note, the primary mechanism of MP release observed in this study was delamination of the embrittled surface, without any macroscopic fracture of the PP specimens. Therefore, the present study focuses on relatively small, μm-scale MPs. In other words, the experimental setup was designed to avoid macroscopic fracture of the specimen, and thus larger, mm-scale MPs generated through such a mode were not observed. It may be possible to produce mm-scale MPs by using sufficiently large specimens and applying extended photo-aging treatment.

Between 0 and 20 h of stirring, MPs were not observed, while the fragmentation of MPs and NPs was observed as an increase in the weight of collected particles (Figure 3) with the presence of a waxy residue. This observation suggests the initial fragmentation of NP (< 1 μm) components. Also, mechanical stress and immersion in water may cause the formation of new cracks [43] and promote crack propagation. This crack propagation may induce MPs (> 1 μm) fragmentation after 20 h of stirring.

Between 20 and 100 h of stirring, MPs (> 1 μm) were fragmented. Compared to Figure 4A, the edges formed by cracks appear more obscure in Figure 4D-4F, implying that MPs may be fragmented through a process in which sharp edges gradually round off. When a mechanical stress is applied to a photo-aged specimen, stress concentration will occur at the crack propagation front. In addition, the oxidation is promoted around cracks [44]. The regions around the edges might be selectively damaged.

In this period, the MP size distribution followed an exponential distribution (Figure 2). No prior reports highlight such a behavior. The reason for this exponential distribution remains unclear, but the fact that an exponential distribution is realized implies the existence of a characteristic length scale which governs the MP fragmentation. One might expect that such a characteristic size is related to hierarchical crystalline structures of PP. However, the spherulite size of PP estimated by a micro-Raman imaging and a polarized optical microscope image was approximately 30 μm (Figure 5). Almost all observed MPs were smaller than the spherulites. Further investigation is required to determine on this issue.

Between 100 and 400 h of stirring, larger MPs (> 10 μm) were fragmented, and the MP size distribution shifted toward power-law distributions (Figure 2). Considering the cutoff imposed by the specimen size itself, the observed power-law behavior may be apparent and valid only within a limited length scale. However, this apparent power-law distribution is consistent with previous reports [11,17,20–26]. To draw definitive conclusions about the statistical characteristics of MP size distributions, more data and a broader investigation across various conditions of MP formation are needed. Nevertheless, the scenario observed in this study—where the MP size distribution transitions from an exponential to a power-law form as stirring progresses—was consistently and robustly observed. Furthermore, we envisage that with sufficient statistical sampling, the exponents of both the exponential and power-law distributions will converge to characteristic values that are independent of stirring time. To our knowledge, no previous study has documented the transition of MP size distribution with increasing stirring time. The observed transition probably reflects a change in the MP fragmentation mechanism. A power-law distribution often arises from some scale-free, fractal processes. Although the detailed mechanism is not clear, it is expected that the MP formation process has some fractal natures. While a detailed discussion of the apparent power-law exponent based solely on the current dataset may not be meaningful, the exponents observed in our results are slightly larger than the commonly reported value of 3. The exponent of 3 is often attributed to the 3D fragmentation model [23], but our results imply that the 3D fragmentation model may not fully explain the MP fragmentation behavior. The origin of the power-law exponent may not be the cascaded fragmentation, but some fractal structures at the surface of photo-aged plastics. For example, propagation pathways of cracks may have a fractal shape. Then the size distribution of MPs fragmented along propagated cracks will reflect the fractal nature. The exponent in the cascaded fragmentation is known to vary and can be sensitive to even simple additional effects [45].

To gain a deeper understanding of the fractal nature and the transition in the functional form of the MP size distribution proposed in this study, it is essential to investigate MP size distributions across a broader size range. In the present study, the observable MP size was inherently limited by the dimensions of the PP specimens used, which imposed an upper cutoff. As part of future work, it will be necessary to design experiments using larger plastic specimens in order to examine the full spectrum of MP sizes released and to better capture the underlying statistical characteristics. In the present study, delamination from the photo-aged embrittled surface is the dominant mode of MP release. However, when dealing with larger systems, it is anticipated that additional (macroscopic) fracture modes may come into play. Whether the trends and interpretations discussed in this work remain valid under such a condition is an also intriguing question.

The MP fragmentation from the photo-aged PP specimen surface is also supported by spectroscopic data (IR and Raman spectra). The IR spectra imply that the chemi-crystallized regions on the surface decrease with stirring. The Raman spectra suggest that the crystallinity of the surface decreases after the fragmentation. Furthermore, using Raman microscopy, we directly compared the local crystallinity of isolated MPs with that of the original PP specimen. The results revealed that the isolated MPs exhibit higher local crystallinity than the stirred PP surface, shedding light on a previously unreported aspect of the degradation and fragmentation process. Although many previous studies have suggested a relationship between bulk chemi-crystallization and crack propagation [15,46], our study provides a novel and direct comparison that has not been reported in the field of polymer degradation chemistry. In addition, these results are consistent with the scenario that chemi-crystallized surface regions are damaged and fragmented into MPs by stirring. Regarding the relationship between MP fragmentation and chemi-crystallization, two contributing factors are considered: (i) stress concentration cracking due to surface embrittlement, and (ii) the presence of amorphous regions surrounding chemi-crystallized areas, where molecular scission has already progressed significantly. It remains unclear which of these factors plays the dominant role. This issue warrants further investigation in future studies. Furthermore, UV irradiation significantly affects the surface region, but the bulk region is not largely affected [47]. At the very long stirring times, the depth dependence of crystallinity may become important.

Small MPs that reflect the material’s molecular-scale nature are difficult to recover once released into the marine environment, yet they likely exert a large impact by providing abundant surface area for adsorption and transport of toxic product and NPs. Our laboratory study, analyzed within the frameworks of degradation chemistry and polymer physics, shows that MP size distributions can be governed by intrinsic molecular-level or higher-order structures. Notably, we find that the relatively small MPs (<10μm) released immediately after photo-aging follow an exponential distribution rather than a power law. Incorporating this exponential early-release regime into MP drift and fate models could improve predictions of coastal and pelagic transport. Accordingly, statistical information on early-release small MPs provides a basis to refine mass-balance estimates (related to "missing ocean plastics"). By linking ocean mechanics with polymer science, environmental forecasting can be improved; the present work represents a step toward that integration.

## 5. CONCLUSION

The time evolution of MP size distribution and the fragmentation rate of MPs and NPs during the stirring process of photo-aged PP specimens were investigated. MPs were fragmented from the photo-aged PP specimen by stirring in deionized water. During the stirring, the size of the fragmented MPs increased with increasing stirring time, and the size distribution transitioned from an exponential to a power-law distribution. Also, the fragmentation rate decreased steeply at the early stage before gradually decreasing at longer stirring times. IR and Raman spectroscopy revealed that MPs were fragments of high crystallinity regions at the specimen surface. The proposed scenario for MP fragmentation is as follows: High crystallinity regions are formed at the surface during the photo-aging process. The surface region becomes brittle and prone to cracking, making it more susceptible to stress concentration-induced fracture. During the stirring, additional cracks are formed and propagated. Eventually, the high crystallinity regions are fragmented along the surface cracks and released as MPs.

**Author Information**

**Corresponding Authors**

Takato Ishida - Department of Materials Physics, Nagoya University, Furo-cho, Chikusa, Nagoya 464-8601, Japan; orcid.org/0000-0003-3919-2348; E-mail: ishida@mp.pse.nagoya-u.ac.jp

**Authors**

Kazuya Haremaki - Department of Materials Physics, Nagoya University, Furo-cho, Chikusa, Nagoya 464-8603, Japan; orcid.org/0009-0007-8494-8008; E-mail: haremaki.kazuya.w2@s.mail.nagoya-u.ac.jp

Takumitsu Kida - Department of Materials Chemistry, Faculty of Engineering, The University of Shiga Prefecture, 2500, Hassaka, Hikone, 522-8533, Japan; orcid.org/0000-0002-9494-3004; E-mail: kida.t@mat.usp.ac.jp

Yusuke Koide - Department of Materials Physics, Nagoya University, Furo-cho, Chikusa, Nagoya 464-8603, Japan; orcid.org/0000-0002-4843-6888; E-mail: koide.yusuke.k1@f.mail.nagoya-u.ac.jp

Takashi Uneyama - Department of Materials Physics, Nagoya University, Furo-cho, Chikusa, Nagoya 464-8603, Japan; orcid.org/0000-0001-6607-537X; E-mail: uneyama@mp.pse.nagoya-u.ac.jp

Yuichi Masubuchi - Department of Materials Physics, Nagoya University, Furo-cho, Chikusa, Nagoya 464-8603, Japan; orcid.org/0000-0002-1306-3823; E-mail: mas@mp.pse.nagoya-u.ac.jp

**Conflict of Interest**

The authors declare no competing interests.

**Acknowledgment**

This work was supported by JSPS KAKENHI Grant Numbers 24K20949, 22KJ1543, "Nagoya University High Performance Computing Research Project for Joint Computational Science" in Japan, CCI holdings Co., Ltd., Mayekawa Houonkai Foundation, Suzuki Foundation, Yazaki Memorial

Foundation for Science and Technology, Fujimori Science and Technology Foundation, Fuji Seal Foundation, The Naito Research Grant and Suga Weathering Technology Foundation, Toukai Foundation for Technology, CASIO Science Promotion Foundation, Hibi Science Foundation, JST PRESTO Grant Number JPMJPR23N3, Japan.

## Supporting Information

# Effects of Stirring Time on Formation of Microplastics Fragmented from Photo-aged Polypropylene

*Kazuya Haremaki[a], Takumitsu Kida[b], Yusuke Koide[a],*

*Takashi Uneyama[a], Yuichi Masubuchi[a], Takato Ishida[a*]*

[a] Department of Materials Physics, Nagoya University, Furo-cho, Chikusa, Nagoya 464-8603, Japan

[b] Department of Materials Chemistry, Faculty of Engineering, The University of Shiga Prefecture, 2500, Hassaka, Hikone, 522-8533, Japan

Tel: +81-52-789-4202, Mobile:+81-90-1145-4372

*E-mail address:* ishida@mp.pse.nagoya-u.ac.jp

TABLE OF CONTENTS

## 1. Reproducibility of MP size distribution across independent runs

We present MP size distributions obtained from three independent experimental runs using different pressed sheets, all prepared following the same procedure described in the manuscript. In the main manuscript, we used the dataset indicated by the black-filled symbols in Figure S1 as a representative example. Due to substantial variability in the degree of photo-aging—originating from the stochastic nature of the induction period—we focused on a single, but typical, dataset for the main analysis. Although some variation is observed among different experimental runs, the transition in the functional form of the MP size distribution—from exponential to power-law—was consistently and robustly observed between 50–100 h and 100–200 h in all cases demonstrated in Figure S1.

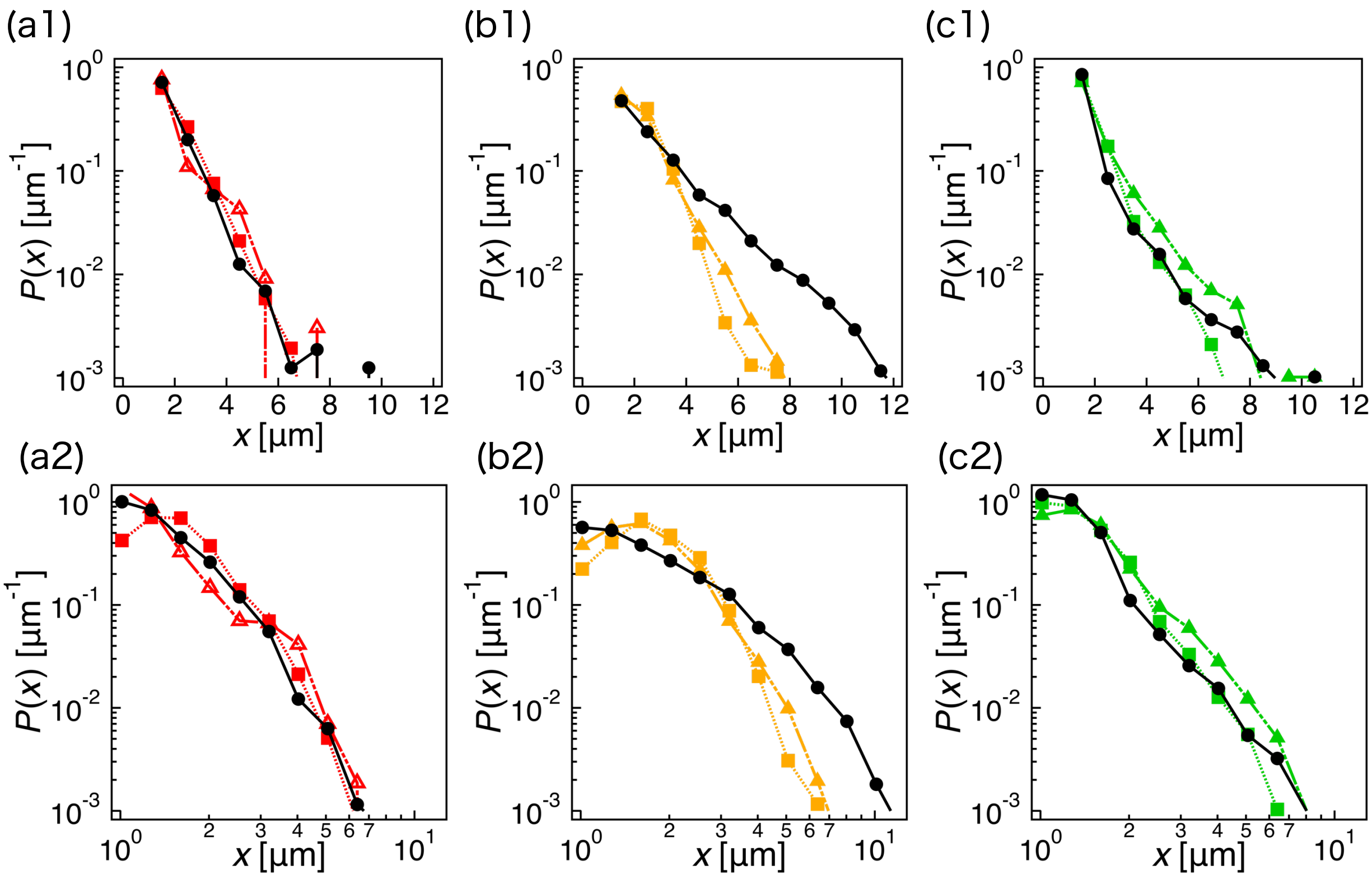

Figure S1 MP size distributions from 3 independent experimental runs, shown for different stirring periods: (a1) semi-logarithmic and (a2) logarithmic plots for 20–50 h; (b1) semi-logarithmic and (b2) logarithmic plots for 50–100 h; and (c1) semi-logarithmic and (c2) logarithmic plots for 100–200 h. The data shown by the black-filled symbols are presented in the manuscript. The bin size is 1 µm for linear and semi-logarithmic scales and logarithmically spaced for the logarithmic scale. In the 20–50 h range, the dataset plotted with hollow triangle symbols contains fewer MPs compared to the others, and its statistical reliability is therefore somewhat lower.

## 2. Reproducibility of Fragmentation Behavior in Independent Experiments

The results from three independent experimental runs for the time evolution of the average fragmentation rate of MPs and NPs, as well as the accumulated weight of fragmented MPs and NPs, are shown in Figure S2. The definitions of $R_{\mathrm{frag},i}$ and $W_{\mathrm{frag},n}$ are identical to those used in the manuscript.

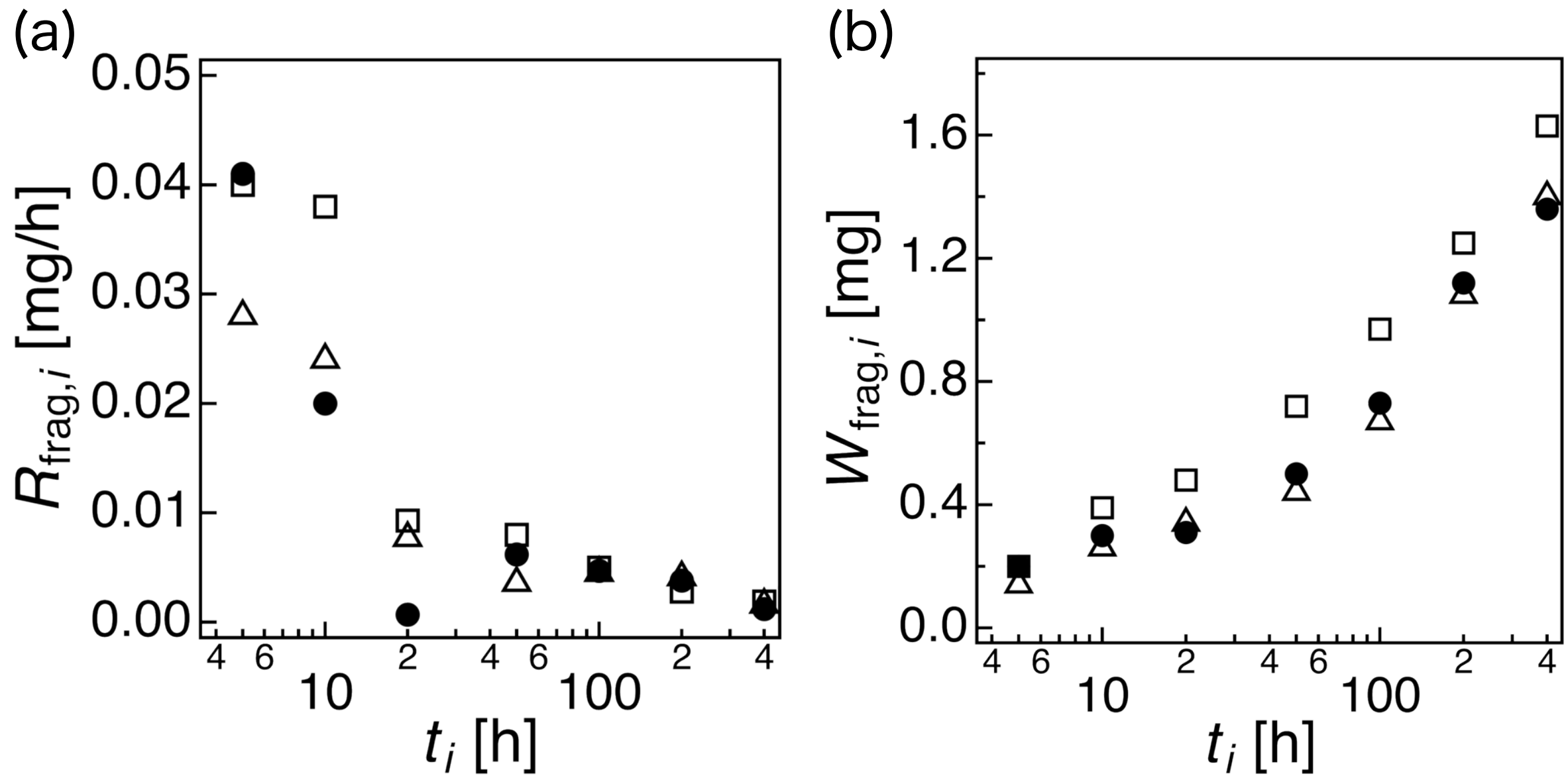

Figure S2 Time evolutions of (a) the average fragmentation rate of MPs and NPs and (b) the accumulated weight of fragmented MPs and NPs, obtained from three independent experimental attempts. $t_i$ is end of the $i$-th stirring period. The data shown by the black-filled symbols are presented in the manuscript.

## 3. Raman spectra analysis

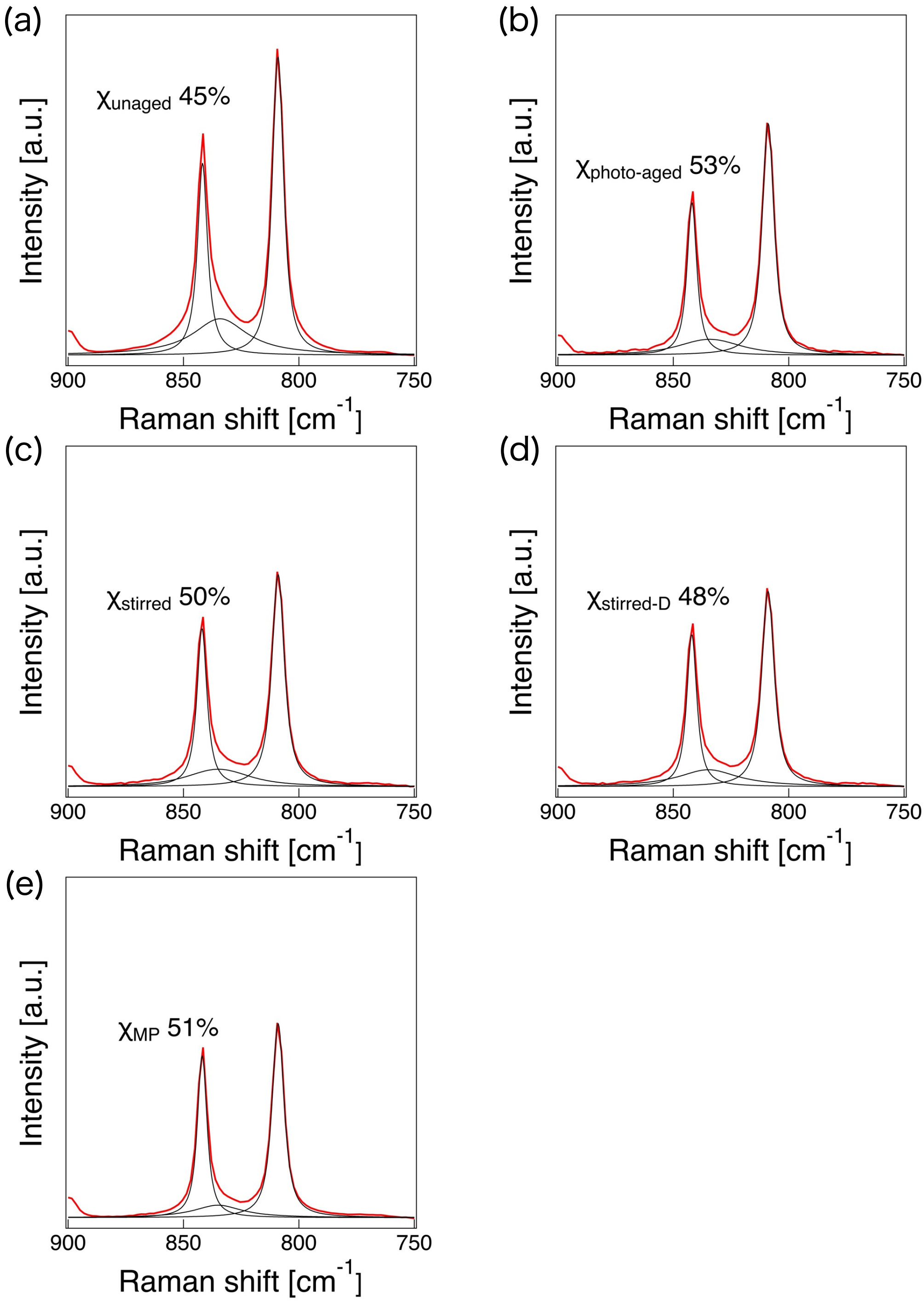

Figure S3 Raman spectra for 900-750 cm$^{-1}$ and peak fitting results of (a) unaged, (b) photo-aged, (c) stirred, (d) stirred-D, (e) MP samples. Spots of the stirred specimen with and without the fragmentation of MPs were referred to as “stirred” and “stirred-D”, respectively. Thin black curves represent the peak fitting results with three Voigt functions.